# Sub-surface Skin Deformation in Response to Gentle Brushing*

Saito Sakaguchi[1], Basil Duvernoy[1], Anders Fridberger[1], Håkan Olausson[1] and Sarah McIntyre[1]

## I. Introduction

There is significant individual variation in tactile function, and the intensity and impression of the same stimulus can differ greatly among individuals [1], [2]. The perception of pleasantness in response to simple stimuli like brushing also varies greatly among individuals [3]. The contributing factors to these large individual variations need to be precisely described in order to be able to develop effective haptic devices and interfaces that can be adapted to individual users.

Tactile nerve endings lie beneath the skin surface and sense mechanical events regulated by the complex structure of the skin tissue. The skin has the potential to regulate sensory perception, including how skin tissue propagates and filters mechanical forces [4], [5]. Efforts to explain the variation in tactile perception using mechanical properties measured from the skin surface are becoming more active [6], [7]. Along these lines, it has been shown that hydrated skin improves the tactile perceptual ability [8], [9], and applying a thin film to the skin reduces the pleasantness of brushing [10]. In parallel, research is being conducted to estimate the mechanical quantities transduced by nerve endings through computer simulations [11], [12]. While most studies still rely on skin surface recordings, there is one investigation of sub-surface skin tissue strains of the fingertip using optical coherence tomography, although this was limited to morphological observations with a low sampling rate [13].

To measure sub-surface skin dynamics, we use a new technology called functional Optical Coherence Tomography (fOCT). The fOCT integrates custom tools, sampling, and analysis methods with a Spectral Domain OCT (SD-OCT) system that enables high refresh rate videos of the skin below the surface [14].

In this study, we focus on the role of the skin, newly recognized not just as a tissue covering the body but as a sensory regulator, and we visualize the internal strains of the skin under tactile stimulation using fOCT. Specifically, we present time-series data of the displacement of the skin caused by brushing stimuli applied to the back of the hand and describe the data acquisition method at different depths of the skin as an example. We chose to investigate a soft brushing stimulus as a pleasant natural stimulus, which is desirable for haptic interfaces to reproduce or emulate. Here we present the method and illustrate its application using data from a single participant.

*Research supported the Swedish Research Council (2024-00381) and Shiseido Co. Ltd.

[1]S. S., B. D., A. F., H. O. and S. M. are with the Department of Biomedical and Clinical Sciences, Linköping University, SE-58183 Linköping, Sweden. e-mail: saito.sakaguchi@liu.se

## II. Method

The experimental setup comprised: (1) a sub-surface imaging system; (2) an automated brushing robot. The imaging system was an OCT system (TEL320C1, Thorlabs, central wavelength 1300 nm), implementing an interferometric imaging technique capable of generating depth-resolved images (around 5.5 microns resolution and 3.5 mm of depth field of view in the air) [15]. Driven with custom-made software, it enabled recordings of sub-surface skin imaging at a refresh rate of 10 kHz. The automated brushing robot combined a collaborative robot "cobot" arm (Universal Robots UR3e) and a force sensor (NRS-6, Nordbo Robotics) with a soft brush attached to it.

The general experimental outline involved simultaneously observing the skin with OCT while brushing the skin on the dorsal surface of the hand (Fig. 1A, B). The force applied by the brush was controlled to be between 0.2 and 0.4 N. Brushing was performed at a constant speed of 3 cm/s in the direction indicated by the arrow in Fig. 1A. These design elements were based on studies showing that many C-tactile fibers (CTs) are found on the back of the hand (innervation zone of the lateral antebrachial nerve), and that the brush's force and speed strongly activate CTs and are perceived to be very pleasant [16], [17]. The OCT continuously observed a single region of interest (ROI) on the skin along the brushing path in a time series at 10 kHz for 2.5 seconds. For data processing, we used morphological and phase data from 5000 time points (0.5 seconds) immediately before and after the moments when the brush blocked the OCT light path for analysis. We first tracked the deformation of the skin surface from the time-series in morphological data (Fig. 1C, blue). Next, we applied a phase-resolved technique [14], [18] to the OCT signal. This technique focuses on the time series of one specific pixel (or specific depth) to generate the frequency domain of the motion. This processing was applied at different depths from the skin surface—0, 38, and 152 pixels, corresponding to 0, 100, and 400 µm—which represent the stratum corneum, epidermis, and dermis layers, respectively.

## III. Results

Fig. 1C shows the time-series data of skin morphology during brushing. The area below the dark blue represents the interior of the skin. The morphology of the skin interior at a depth of approximately 500 µm (200 pixels) could be observed. Fig. 1D shows the FFT analysis results of phase change data inside the skin before and after the brush passes through the ROI of the OCT. Both before and after brushing, there was a large amplitude in the low-frequency region below 30 Hz. After brushing, a peak in amplitude was also observed around 60 Hz. Furthermore, at this peak, there were different peaks at each depth, with the interior of the skin (orange, green) showing larger peaks than the skin surface (blue).

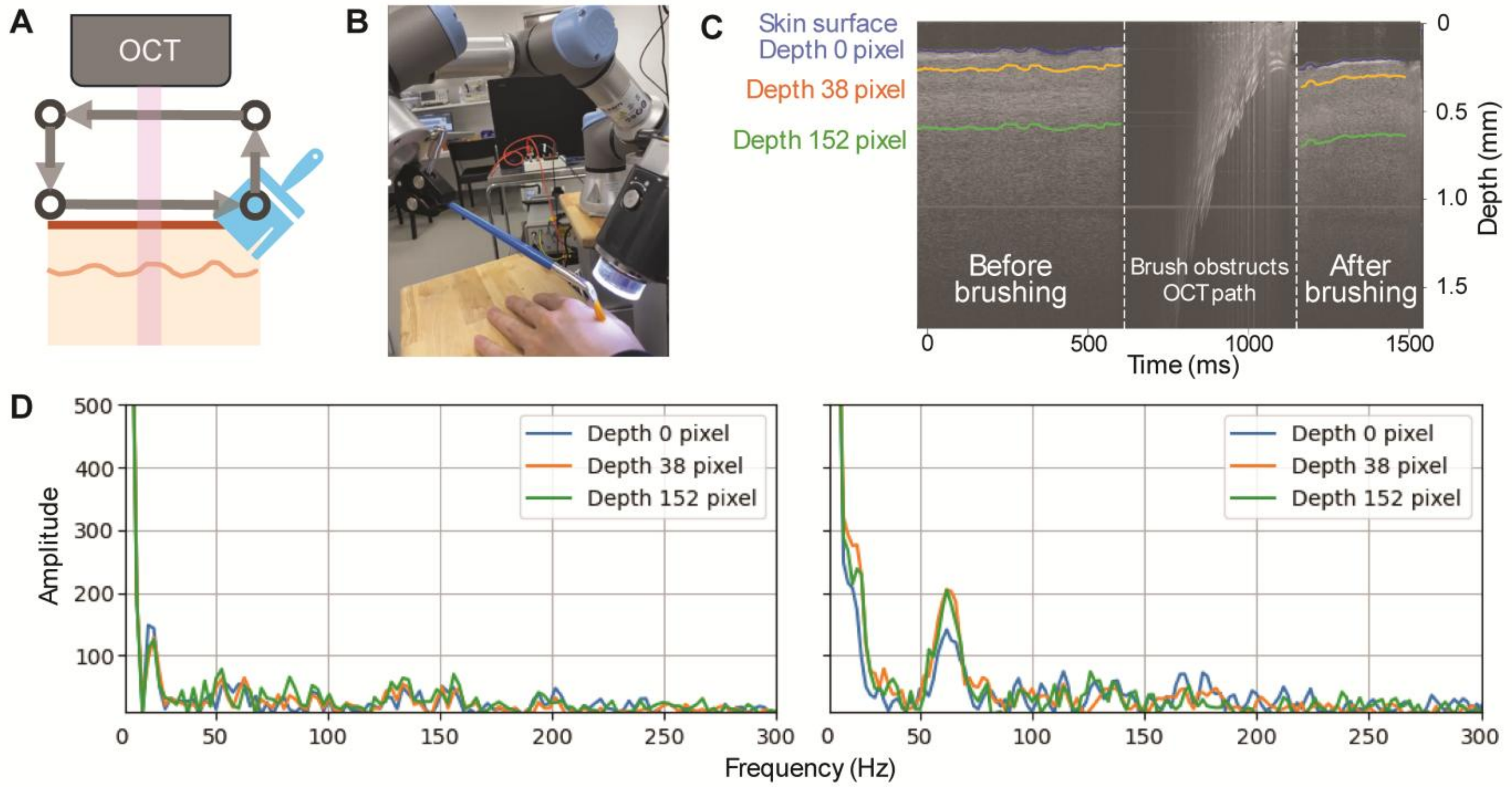

Figure 1. Experimental overview and representative OCT data. *(A: Schematic of OCT observation during brushing. B: Actual experimental setup. C: Time-series data of skin morphology at a single ROI. D: Frequency analysis of pixel-wise phase changes before (left) and after (right) brushing the ROI)*

## IV. Discussion

We found that using OCT phase data allows us to quantify minute displacement with a temporal resolution of 10 kHz. The frequency analysis graphs before and after brushing may reflect the vibration and compression of the skin caused by the brush approaching the ROI in the OCT, as well as the extension of the skin caused by the brush passing through the ROI. The different displacement patterns observed at different depths can be attributed to the heterogeneous and complex layered structure and viscoelastic properties of the skin. This phenomenon may share the same structural mechanism as the findings that deeper tissue absorb most of the fingertip's lateral movement rather than the surface layer [13]. Brushing in the tangential direction may cause different strains in each skin layer due to the complex interactions between the surface layer and deeper tissues. Future research, by increasing the variety of stimuli and intervening skin properties, could lead to a deeper understanding of the contribution of skin dynamics to tactile perception.